\documentclass{iau} 
\usepackage{graphicx}

\title[VISTA of the Small Magellanic Cloud] 
{A near-infrared VISTA of the\\ Small Magellanic Cloud}

\author[Maria-Rosa L. Cioni,  et al.]   
{Maria-Rosa L. Cioni$^1$, Florian Niederhofer$^1$, Stefano Rubele$^{2,3}$, Ning-Chen Sun$^{4, 5}$, 
  }

\affiliation{
$^1$Leibniz Institut f\"{u}r Astrophysik Potsdam, An der Sternwarte 16, D-14482 Potsdam, Germany\\ email: {\tt mcioni@aip.de}\\[\affilskip]
$^2$INAF-Osservatorio Astronomico di Padova, vicolo dell'Osservatorio 5, I-35122, Padova, Italy\\ email: {\tt stefano.rubele@inaf.it}\\[\affilskip]
$^3$Dipartimento di Fisica e Astronomia, Universit\`{a} di Padova, vicolo dell'Osservatorio 2, I-35122, Padova, Italy\\[\affilskip]
$^4$Department of Physics and Astronomy, University of Sheffield, Hicks Building, Houinsfield Road, Sheffield S3 7RH, United Kingdom\\ email:{\tt sunnc@foxmail.com}\\[\affilskip]
$^5$Kavli Institute for Astronomy \& Astrophysics and Department of Astronomy, Peking University, Yi He Yuan Lu 5, Hai Dian District, Beijing 100871, People's Republic of China 
}

\pubyear{2015}
\volume{344}  
\jname{Dwarf Galaxies: from the Deep Universe to the Present}
\editors{S. Stierwaldt  \& K. B. McQuinn, eds.}
\begin{document}

\maketitle

\begin{abstract}
VISTA observed the Small Magellanic Cloud (SMC), as part of the VISTA survey of the Magellanic Clouds system (VMC), for six years (2010-2016). The acquired multi-epoch $YJK_\mathrm{s}$ images have allowed us to probe the stellar populations to an exceptional level of detail across an unprecedented wide area in the near-infrared. This contribution highlights the most recent VMC results obtained on the SMC focusing, in particular, on the clustering of young stellar populations, on the proper motion of stars in the main body of the galaxy and on the spatial distribution of the star formation history.

\end{abstract}

\firstsection 
\section{Introduction}
The VISTA survey of the Magellanic Clouds system (VMC; \cite[Cioni et al. 2011]{cioni11}) is a near-infrared deep and wide-field study of the history and structure of the Magellanic Clouds. Observations are performed with the VIRCAM instrument mounted at the VISTA telescope, they began in late 2009 and are about to be completed. The VMC survey covers a total sky area of $\sim170\deg^2$ with a spatial resolution of $<1^{\prime\prime}$ in the $YJK_\mathrm{s}$ filters reaching a sensitivity of about $21$ mag (S/N$=10$ in the Vega system). The VMC survey comprises of multi-epoch observations, at least $3$ epochs in the $Y$ and $J$ filters and at least $12$ epochs in the $K_\mathrm{s}$ filter. More than thirty refereed journal papers have been published by the VMC survey team to-date and in this contribution we highlight the results from the three most recent ones.  

\section{Hierarchical star formation in the SMC}
\begin{figure}
\begin{center}
 \includegraphics[width=10cm]{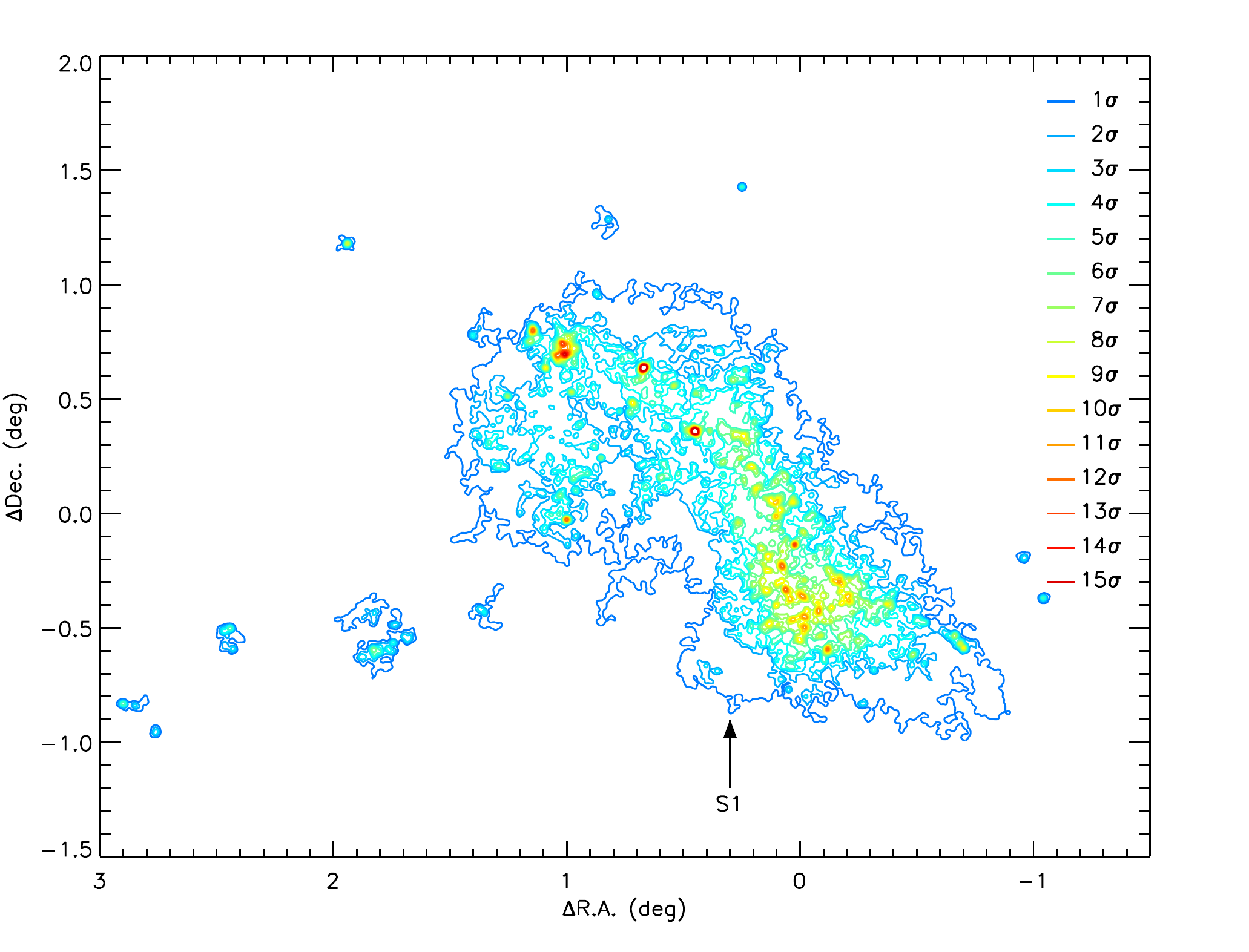} 
 \caption{Boundaries of all identified young stellar structures, with the colours coded according to their significance levels, see \cite[Sun et al. (2018)]{sun18} for details. The map is centred at  ($\alpha_{2000}=00^\mathrm{h}$:$50^\mathrm{m}$:$20.8^\mathrm{s}$, $\delta_{2000}=-72^\circ$:$49^\prime$:$43^{\prime\prime}$). Note that the gap in VMC observations (cf.~Fig.~\ref{fig2}) was filled using data from the Magellanic Clouds Photometric Survey (MCPS; \cite[Zaritsky et al. 2000]{zaritsky2002}).}
   \label{fig1}
\end{center}
\end{figure}

The distribution of young ($<250$ Myr old) upper main sequence stars across the SMC was studied using a cluster analysis technique in order to understand their formation mechanism (\cite[Sun et al. 2018]{Sun18}). Stars were selected from the VMC colour-magnitude diagram, $J-K_\mathrm{s}$ versus $K_\mathrm{s}$, to lie within a box on the upper main sequence, while taking into account of possible shifts due to metallicity, extinction, and the SMC's line-of-sight depth. They were used to identify $556$ separate over densities on $15$ significance levels (Fig.~\ref{fig1}). These structures appeared to follow a hierarchical distribution such that larger structures on lower significance levels contain smaller ones on higher levels. Their properties (size, number of stars, and surface density) as well as their irregular morphologies (characterised by a perimeter$-$area dimension of $D_\mathrm{p}=1.44\pm0.02$) are similar to those of the interstellar medium which is regulated by supersonic turbulence. This highlights the importance of turbulence in controlling star formation over a large scale range.

\section{The proper motion of the SMC}
\begin{figure}
\begin{center}
 \includegraphics[width=12cm]{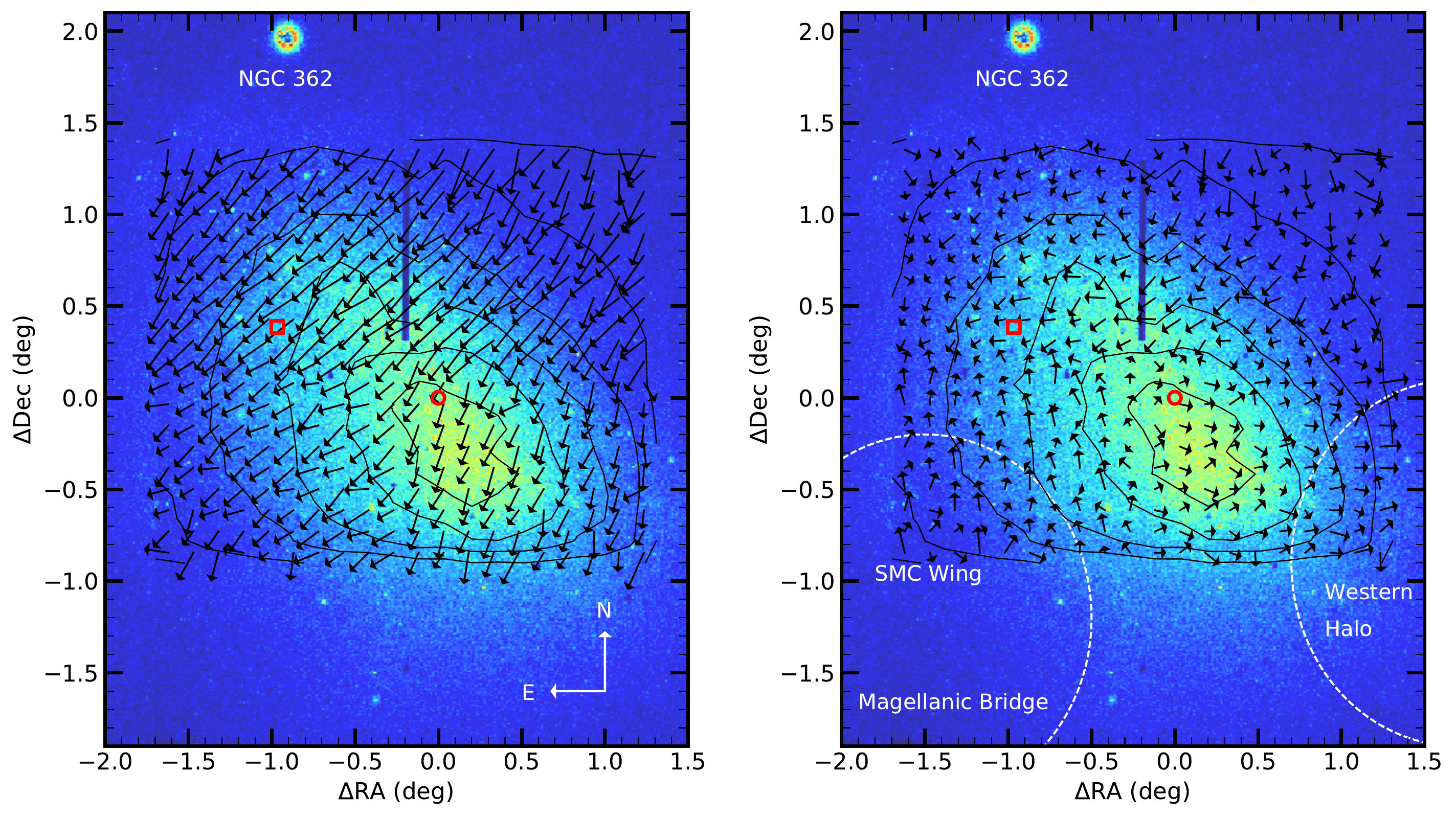} 
 \caption{
 Proper-motion field (black arrows) of the central regions of the SMC from four tiles of VMC data superimposed to a star density map. Each panel is centred at the optical centre of the SMC (red circle; $\alpha_{2000}=00^\mathrm{h}$:$52^\mathrm{m}$:$12.5^\mathrm{s}$, $\delta_{2000}=-72^\circ$:$49^\prime$:$43^{\prime\prime}$), also shown as a red square is the kinematical H I centre. North of the main body of the SMC, the foreground Galactic globular cluster NGC 362 is visible. The contours are at 500, 1500, 3500, 5500, 7500 and 9500 stars per grid cell. The vertical dark stripe at ÆRA$=-0.2^\circ$ is due to a narrow gap in the observations. In the left-hand panel, the arrows indicate the observed absolute proper motion, whereas in the right-hand panel, the arrows show the residual proper motions after subtraction of the systemic velocity of the SMC, see \cite[Niederhofer et al. (2018)] {niederhofer18} for details.
 }
   \label{fig2}
\end{center}
\end{figure}

The proper motions of the stellar populations in the centre of the SMC have been derived using multi-epoch VMC observations. Four VMC tiles were examined and the reflex motions of about $33,000$ background galaxies was measured with respect to about $700,000$ stars. The proper motions were measured from the pixel displacements in the x and y directions of the source centroids which were obtained from point-spread-function photometry on individual pawprint  $K_\mathrm{s}$-band detector images. On average, $12$ independent epochs over a time baseline of $\sim24$ months were available and each detector contained about $6,000$ stars and $150$ background galaxies. A median systemic proper motion of the SMC,  $(\mu_\alpha cos(\delta), \mu_\delta)=(1.087, -1.187)$ mas yr$^{-1}$ with systematic uncertainties of $0.192$ and $0.008$ mas yr$^{-1}$ for the two components and a random uncertainty of $0.003$ mas yr$^{-1}$, was found. This is consistent with previous studies. Furthermore, from the median-subtracted proper motion distribution (Fig.~\ref{fig2}) we identified regions of non-uniform velocity, in the densest area and in the southeast of the galaxy, probably associated to tidal stripping events. While the latter feature points towards the Wing and the Bridge, the former might be associated to the Counter-Bridge (\cite[Diaz \& Bekki 2012]{Diazandbekki12}).

\section{The star formation history of the SMC}
\begin{figure}
\begin{center}
\resizebox{0.33\hsize}{!}{ \includegraphics{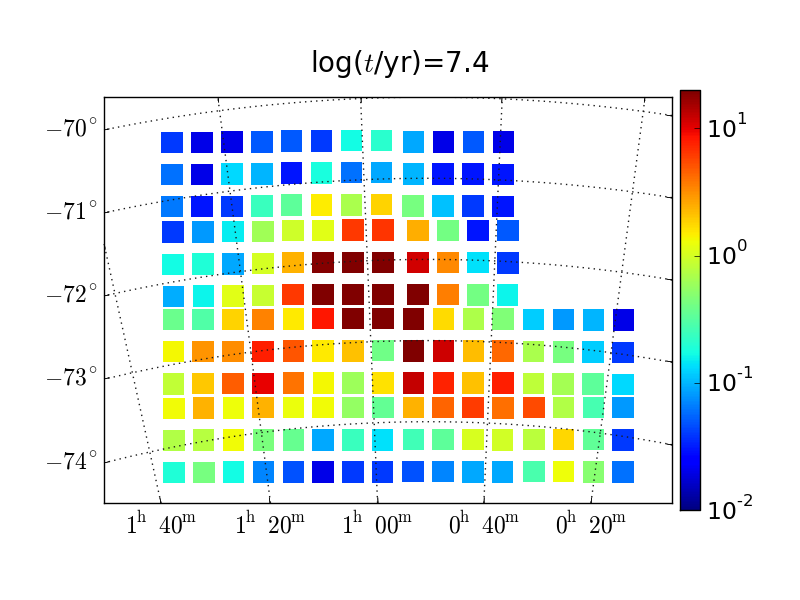}}
\resizebox{0.33\hsize}{!}{ \includegraphics{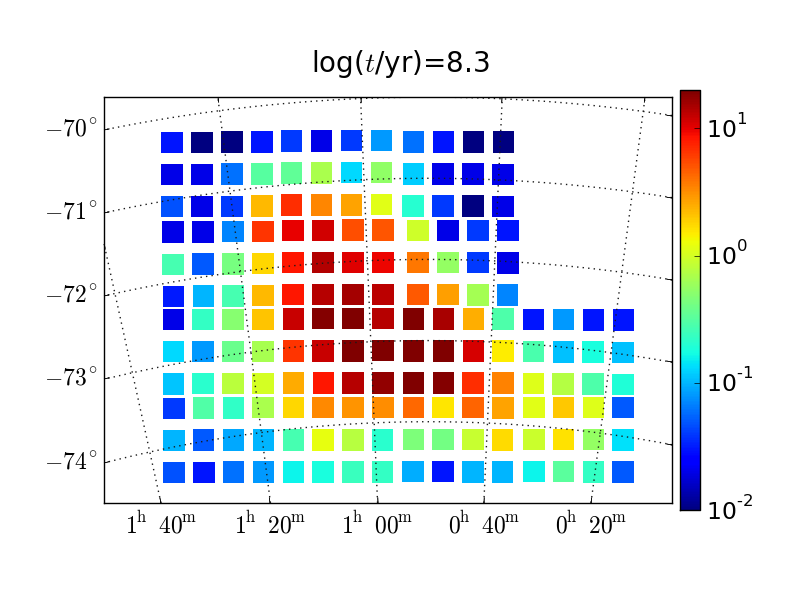}}
\resizebox{0.33\hsize}{!}{ \includegraphics{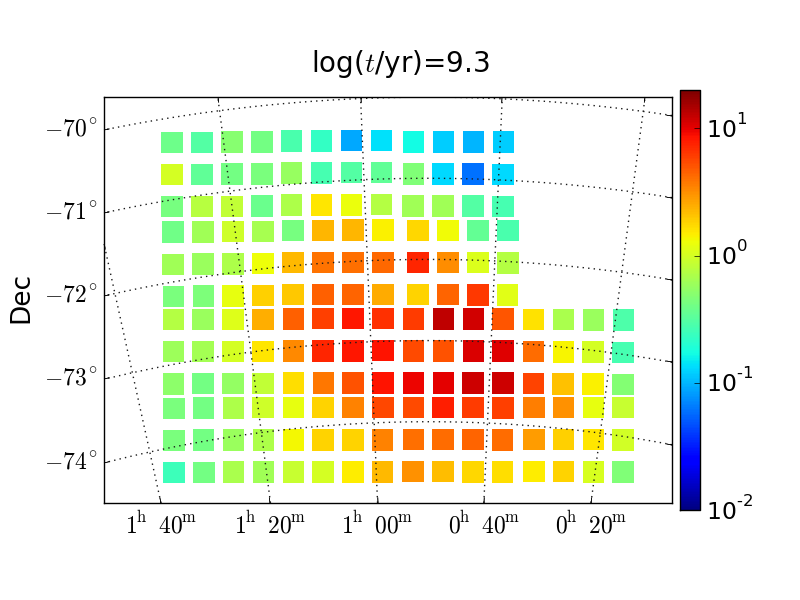}}
\resizebox{0.33\hsize}{!}{ \includegraphics{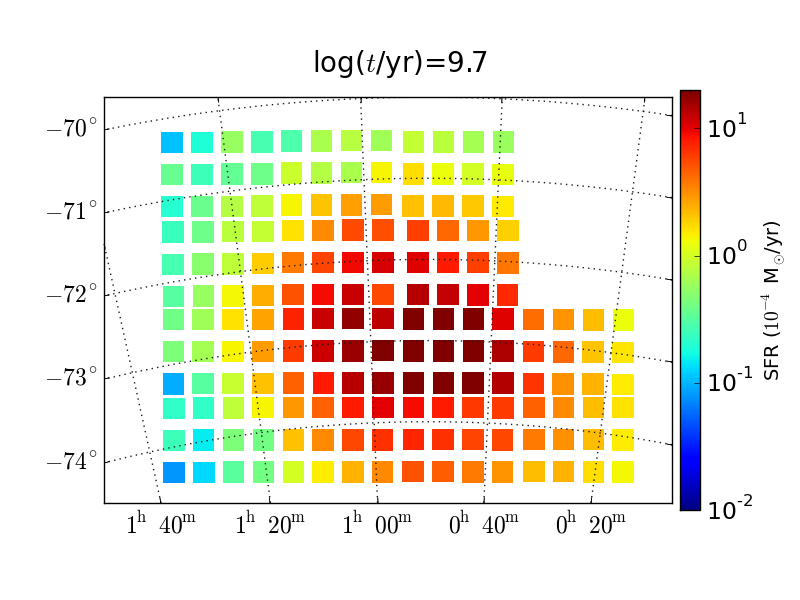}}
 \caption{
 Star formation rate maps across the SMC as a function of age, for four out of fourteen age bins considered in \cite[Rubele et al. (2018)]{rubele18}. Each square corresponds to a subregion of $0.143$ deg$^2$ in size, with colours indicating the star formation rate in units of $10^{-4}$M$_\odot$ yr$^{-1}$.}
   \label{fig3}
\end{center}
\end{figure}

The star formation history across the central $\sim24$ deg$^2$ region of the SMC ($14$ VMC tiles) has been derived from the reconstruction of the observed VMC colour-magnitude diagrams ($Y-K_\mathrm{s}$ versus $K_\mathrm{s}$ and $J-K_\mathrm{s}$ versus $K_\mathrm{s}$) by  \cite[Rubele et al. (2018)]{rubele18}. The analysis was performed within individual subregions of $0.143$ deg$^2$ in size. Figure \ref{fig3} shows that at young ages ($log(t)=7.4$ or $\sim12$ Myr) the star formation is concentrated in the Bar and Wing areas while the separation between these two substructures is less clear at older ages. A northwestern edge, possibly resulting from the dynamical interaction with the Large Magellanic Cloud ($log(t)=8.3$ or $\sim200$ Myr ago), becomes well delineated. The distribution traced by old ($log(t)=9.7$ or $\sim 5$ Gyr old) stars is significantly rounder than that resulting from younger stars. We explored the age interval from $8$ Myr to $10$ Gyr and all maps are included in \cite[Rubele et al. (2018)]{rubele18}. In this study, we also derived the spatial distribution of the mean reddening and distance of the stellar populations as well as an estimate of the total stellar mass, $(5.31\pm0.05)\times10^8$ M$_\odot$, that the galaxy produced over its lifetime.

\section{Conclusions}
The VMC observations across the SMC area were completed in 2016 and the data are currently being reprocessed with the latest version of the CASU pipeline (v1.5) to adjust the zero points and to apply a correction to the distortion of tile images, more details are given in \cite[Gonzalez-Fernandez et al. (2018)]{Gonzalez18}. Several works have resulted from the previous reductions (i.e.~v1.3) and analysis of the data, e.g.: a study of the three-dimensional structure of the SMC using Cepheids (\cite[Ripepi et al. 2017]{Ripepi17}) and RR Lyrae stars (\cite[Muraveva et al. 2018]{Muraveva18}), the discovery of a bi-modality in the red-clump distribution (\cite[Subramanian et al. 2017]{Subramanian17}) and of new star clusters (\cite[Piatti et al. 2016]{Piatti16}), while in this contribution we focused on the three most recent publications. In future works, we will complement the star formation history and the proper motion analysis of the SMC with VMC tiles covering the external regions of the galaxy. We also plan to extend these studies, the study of the clustering of young stellar structures and that of the three-dimensional structure to the Large Magellanic Cloud.


\begin{thebibliography}{}

\bibitem[Cioni \etal\ (2011)]{Cioni2011}
{Cioni, M.-R. L., Clementini, G., Girardi, L., et al.} 2011, 
\textit{A\&A}, 527, A116

\bibitem[Diaz \& Bekki (2012)]{Diazandbekki12}
{Diaz, J. D., \& Bekki, K.} 2012, 
\textit{ApJ}, 750, 36

\bibitem[Gonzalez-Fernandez \etal\ (2018)]{Gonzalez18}
{Gonzalez-Fernandez, C., Hodgkin, S. T., Irwin, M. J., et al.} 2018, 
\textit{MNRAS}, 474, 5459

\bibitem[Muraveva \etal\ (2018)]{Muraveva18}
{Muraveva, T., Subramanian, S., Clementini, G., et al.} 2018,
\textit{MNRAS}, 473, 3131

\bibitem[Niederhofer  \etal\  (2018)]{Niederhofer18}
{Niederhofer, F., Cioni, M.-R. L., Rubele, S., et al.} 2018,
\textit{A\&A}, 613, L8

\bibitem[Piatti \etal\ (2016)]{Piatti16}
{Piatti, A. E., Ivanov, V. D., Rubele, S., et al.} 2016,
\textit{MNRAS}, 460, 383

\bibitem[Ripepi \etal\ (2017)]{Ripepi17}
{Ripepi, V., Cioni, M.-R. L., Moretti, M. I., et al.} 2017,
\textit{MNRAS}, 472, 808

\bibitem[Rubele \etal\ (2018)]{rubele18}
{Rubele, S., Pastorelli, G., Girardi, L., et al.} 2018,
\textit{MNRAS}, 478, 5017

\bibitem[Subramanian  \etal\ (2017)]{Subramanian17}
{Subramanian, S., Rubele, S., Sun, N.-C., et al.} 2017,
\textit{MNRAS}, 467, 2980

\bibitem[Sun \etal\ (2018)]{sun18}
{Sun, N.-C., de Grijs, R., Cioni, M.-R. L., et al.} 2018,
\textit{ApJ}, 858, 31

\bibitem[Zaritsky \etal\ (2002)]{zaritsky02}
{Zaritsky, D., Harris, J., Thompson, I. B., et al.} 2002,
\textit{AJ}, 123, 855

\end{thebibliography}
\end{document}